\documentclass[12pt,twoside]{article}

\def\TheAuthor{Michael Bachmann}                                   
\def\TheTitle{Monte Carlo Simulations}                        
\def\TheHeading{Monte Carlo Simulations}                                      
\def\TheInstitute{Institut f\"ur Festk\"orperforschung (IFF-2) and Institute for Advanced Simulation (IAS-2)}       
\def\TheAddress{Forschungszentrum J\"ulich, D-52425 J\"ulich, Germany}              
\def\ThePart{A}                                               
\def\TheChapter{10}                                            

\usepackage{ferienschule_11}                                  

\begin{document}

\MakeTitel           

\tableofcontents     

\newpage


\section{Introduction}
\index{ensemble!microcanonical}\index{ensemble!canonical}
\index{Monte Carlo simulation}
For a system under thermal conditions in a heat bath with temperature $T$,
the dynamics of each of the system particles is influenced by
interactions with the heat-bath particles. If quantum effects are negligible
(what we will assume in the following), the classical motion of any
system particle looks erratic; the particle follows a stochastic path. The system can ``gain''
energy from the heat bath by these collisions (which are typically more
generally called ``thermal fluctuations'') or ``lose'' energy by friction
effects (dissipation). The total energy of the coupled system of heat
bath and particles is a conserved quantity, i.e., fluctuation and dissipation refer to
the energetic exchange between heat bath and system particles only. Consequently, the
coupled system is represented by a \emph{microcanonical ensemble}, whereas the
particle system is in this case represented by a \emph{canonical ensemble}: The
energy of the particle system is not a constant of motion. Provided heat bath and system are in thermal
equilibrium, i.e., heat-bath and system temperature are identical, fluctuations
and dissipation balance each other. This is the essence of the celebrated
fluctuation-dissipation theorem~\cite{kubo1}. In equilibrium, only the statistical mean of the
particle system energy is constant in time.

This \emph{canonical} behavior of the system particles is not accounted for by
standard Newtonian dynamics (where the system energy is considered to be a constant of
motion). In order to perform molecular dynamics (MD) simulations of the system
under the influence of thermal fluctuations, the
coupling of the system to the heat bath is required. This is provided by
a thermostat, i.e., by extending the equations of motion by additional heat-bath coupling
degrees of freedom~\cite{frenkelsmith1}. The introduction of thermostats into
the dynamics is a notorious problem in MD
and it cannot be considered to be solved satisfactorily to date~\cite{schbj1}. In order to take
into consideration the stochastic nature of any particle trajectory in the heat bath,
a typical approach is to introduce random forces into the dynamics. These
forces represent the collisions of system and heat-bath particles on the basis
of the fluctuation-dissipation theorem~\cite{kubo1}.

Unfortunately, MD simulations of complex systems on microscopic and mesoscopic
scales are extremely slow, even on the largest available computers. A prominent
example is the folding of proteins\index{protein folding} with natural time scales of milliseconds to
seconds. It is currently still impossible to simulate folding events of bioproteins under
realistic conditions, since even longest MD runs are hardly capable of
generating single trajectories of more than a few
microseconds.  
Consequently, if the intrinsic time scale of a realistic model exceeds the time
scale of an MD simulation of this model, MD cannot seriously be applied in these cases.

However, many interesting questions do not require to consider the intrinsic dynamics of
the system explicitly. This regards, e.g., equilibrium thermodynamics, which includes
all relevant phenomena of cooperativity\index{cooperativity} -- the collective original source for
the occurrence of phase transitions\index{phase transition}. Stability of all matter, independently
whether soft or solid, requires fundamental ordering principles. We are far away
from having understood the \emph{general physical properties} of transition processes that separate, e.g., ordered
and disordered phases, crystals and liquids, glassy and globular polymers,
native and intermediate protein
folds, ferromagnetic and paramagnetic states of metals, Bose-Einstein
condensates and bosonic gases, etc. Meanwhile, the history of research of
collective or critical phenomena has already lasted for more than hundred years and
the universality hypothesis has already been known for several decades~\cite{kadanoff1}. Though,
no complete theory exists which is capable relating to each other phenomena such as protein
folding (unfolding) and freezing (melting) of solid matter. The reason is that
the first process is dominated by finite-size effects, whereas the latter seems
to be a macroscopic ``bulk'' phenomenon. However, although doubtlessly
associated to different length scales which differ by orders of magnitude, both
examples are based on cooperativity, i.e., the collective multi-body interplay of a
large number of atoms. Precise theoretical analyses are extremely difficult,
even more, if several attractive and repulsive interactions compete with each
other and if the system does not possess any obvious internal symmetries (which
is particularly apparent for ``glassy'' heteropolymers like proteins). On the experimental
side, the situation has not been much better as the resolution of the data often
did not allow an in-depth analysis of the simultaneous microscopic effects
accompanying cooperative phenomena. This has dramatically improved by
novel experimental techniques enabling to measure the response of the system to
local manipulations, giving insight in the mesoscopic and macroscopic multi-body effects upon activation. On
the other hand, a systematic understanding requires a theoretical basis.
The relevant physical forces have been known for a long time, but the
efficient combination of technical
and algorithmic prerequisites has been missing until recently. The general
understanding of cooperativity in complex systems as a statistical effect,
governed by a multitude of forces acting on
different energy and length scales, requires the study of the interplay of entropy\index{entropy}
and energy. The key to this is currently only provided by Monte Carlo
computer simulations~\cite{binderLandau1}.

\section{Conventional Markov-chain Monte Carlo sampling}

\subsection{Ergodicity and finite time series}
\index{ergodicity}
The general idea behind all Monte Carlo methodologies is to provide an efficient stochastic sampling
of the configurational or conformational phase space or parts of it with the objective to
obtain reasonable approximations for statistical quantities such as expectation values,
probabilities, fluctuations, correlation functions, densities of states, etc. 

A given system conformation
(e.g., the geometric structure of a molecule) ${\bf X}$ is locally or globally modified 
to yield a conformation ${\bf X}'$. This update or ``move'' is then accepted with
the transition probability $t({\bf X}\to {\bf X}')$. Frequently used updates
for polymer models are, for example, random translational changes of single monomer positions,
bond angle modifications, or rotations about covalent bond axes. More global updates
consist of combined local updates, which can be necessary to satisfy constraints such as fixed 
bond lengths or simply to improve efficiency. It is, however, a necessary condition for
correct statistical sampling that
Monte Carlo moves
are ergodic, i.e., the chosen set of moves must, in principle,
guarantee to reach any conformation out of any other conformation. Since this is often hard
to prove and an insufficient choice of move sets can result in systematic errors, great care 
must be dedicated to choose appropriate moves or sets of moves. Since molecular models often
contain constraints, the construction of global moves can be 
demanding. Therefore, reasonable and efficient moves have to be chosen in correspondence to 
the model of a system to be simulated. 

A Monte Carlo update 
corresponds to the discrete ``time step'' $\Delta \tau_0$ 
in the simulation process. In order to reduce correlations, typically a number
of updates is performed between measurements of a quantity $O$. 
This series of updates is called a ``sweep''
and the ``time'' passed in a single sweep is $\Delta\tau = N\Delta \tau_0$ if the sweep consists of $N$ updates.
Thus, if $M$ sweeps are performed, the discrete
``time series'' is expressed by the vector 
$(O(\tau_\text{init}+\Delta\tau),O(\tau_\text{init}+2\Delta\tau),\ldots,O(\tau_\text{init}+m\Delta\tau),\ldots,O(\tau_\text{init}+M\Delta\tau))$
and represents the Monte Carlo trajectory. The period of equilibration $\tau_\text{init}$ sets the
starting point of the measurement. For convenience, we  
use the abbreviation $O_m\equiv O(\tau_\text{init}+m\Delta\tau)$ and
$\tau_m=\tau_\text{init}+m\Delta\tau$ with $m=1,2,\ldots,M$ in the following. 

According to the 
theory of ergodicity, averaging a quantity over an infinitely long time series is identical to perform
the statistical ensemble average: 
\begin{equation}
\label{eq:mperm:erg}
\overline{O}=\lim_{M\to\infty} \frac{1}{M}\sum\limits_{m=1}^{M} O_m \equiv \langle O\rangle =
\int {\cal D}X O(\textbf{X}) p(\textbf{X}),
\end{equation}
where ${\cal D} X$ represents the formal integral measure for the infinitesimal scan
of the conformation space and $p({\bf X})$ is the energy dependent microstate
probability of the conformation $\textbf{X}$ in the relevant
ensemble in thermodynamic equilibrium [in the canonical ensemble with
temperature $T$, simply $p({\bf X})=\exp[-E({\bf X})/k_\text{B}T]$].
This is the formal basis for Monte Carlo sampling. However, only finite time series can be simulated 
on a computer. For a finite number of sweeps $M$ in a sample $k$, 
the relation~(\ref{eq:mperm:erg}) can only be
satisfied approximately, $M^{-1}\sum_{m=1}^M O^{(k)}_m=\overline{O}^{(k)}\approx \langle O\rangle$. 
Note that the mean value $\overline{O}^{(k)}$ will
depend on the sample $k$, meaning that it is likely that another sample $k'$ will yield
a different value $\overline{O}^{(k')}\neq \overline{O}^{(k)}$. In order to define a reasonable
estimate for the statistical error, it is necessary to start from the assumption 
that we have generated an infinite
number of independent samples $k$. In this case the distribution of the estimates 
$\overline{O}^{(k)}$ is Gaussian, according to the central limit theorem of
uncorrelated samples. 
The exact average of the estimates is then given by $\langle \overline{O}\rangle$.
The statistical error of $\overline{O}$ is thus suitably defined as the standard deviation
of the Gaussian:
\begin{equation}
\label{eq:mperm:errest}
\varepsilon_{\overline{O}}=
\sqrt{\left\langle \left(\overline{O}-\langle \overline{O}\rangle \right)^2\right\rangle}=
\sqrt{\langle \overline{O}^2\rangle-\langle \overline{O}\rangle^2}
=\sqrt{\frac{1}{M^2}\sum\limits_{m=1}^M\sum\limits_{n=1}^M A_{mn}\sigma^2_{O_m}},
\end{equation}
where
\begin{equation}
\label{eq:mperm:acf}
A_{mn}=\frac{\langle O_mO_n\rangle- \langle O_m\rangle\langle O_n\rangle}{\langle O_m^2\rangle- \langle O_m\rangle^2}
\end{equation}
is the autocorrelation function\index{autocorrelation!function} and 
$\sigma^2_{O_m}=\langle O_m^2\rangle-\langle O_m\rangle^2$ is the variance of the 
distribution of individual data $O_m$.   
If the Monte Carlo
updates in each sample are performed completely randomly without memory, i.e., a new conformation is
created independently of the one in the step before (which is a possible but typically 
very inefficient strategy), two measured values $O_m$ and $O_n$ are uncorrelated, if $m\neq n$. Then, the
autocorrelation function
simplifies to $A_{mn}=\delta_{mn}$ and the statistical error satisfies the celebrated relation
\begin{equation}
\label{eq:mperm:errestuncorr}
\varepsilon_{\overline{O}}=\frac{\sigma_{O_m}}{\sqrt{M}}.
\end{equation}
Since the exact distribution of $O_m$ values and the ``true'' expectation value $\langle O\rangle$ 
are unchanged
in the simulation (but unfortunately unknown), the standard deviation $\sigma_{O_m}$ is constant, too. Thus,
the statistical error decreases with $1/\sqrt{M}$.\footnote{For the actual calculation, it is a problem
that $\sigma^2_{O_m}$ is unknown. However, what can be estimated is 
$\tilde\sigma^2_{O_m}=\overline{O^2}-\overline{O}^2$ and for its expected value we thus obtain
$\langle\tilde\sigma^2_{O_m} \rangle=\sigma^2_{O_m}(1-1/M)$. The $1/M$ correction is the
\emph{systematic} error due to the finiteness of the time series, called bias. 
The bias-corrected relation for
the statistical error reads finally $\varepsilon_{\overline{O}}=[M(M-1)]^{-1/2}\sqrt{\sum_m(O_m-\overline{O})^2}$~\cite{wj2002}.} 

In practice, most of the efficient Monte Carlo techniques
generate correlated
data, in which case we have to fall back to the more 
general formula~(\ref{eq:mperm:errest}). It can conveniently be rewritten
as 
\begin{equation}
\label{eq:mperm:errcorr}
\varepsilon_{\overline{O}}=\sigma_{O_m}/{\sqrt{M_\text{eff}}}
\end{equation}
with the effective 
statistics $M_\text{eff}=M/\Delta\tau_\text{ac}\le M$, where $\Delta\tau_\text{ac}$ corresponds to 
the autocorrelation time.\index{autocorrelation!time} This means, the statistics is
effectively reduced by the number of 
sweeps until the correlations have decayed.\footnote{For a detailed discussion of the 
autocorrelation function and the calculation of the
autocorrelation time, see, e.g., Ref.~\cite{wj2002}.} Since it takes at least the time
$\Delta\tau_\text{ac}=N_\text{ac}\Delta \tau_0$ to generate statistically
independent conformations, a sweep can simply contain as many updates $N_{\rm ac}$ as necessary to satisfy 
$\Delta\tau \approx \Delta\tau_\text{ac}$ without losing effective statistics. 
In this case, the $M \approx M_\text{eff}$ 
data entering into the effective statistics are virtually uncorrelated. This is also the general
idea behind advanced, computationally convenient 
error estimation methods such as binning and jackknife
analyses~\cite{jack,wj2002}.
For the correctness of the measurements, $M \approx M_\text{eff}$ is not a
necessary condition; more sweeps with less updates in each sweep, i.e., periods
between  
measurements shorter than $\Delta\tau_\text{ac}$ only yield redundant 
statistical information. This is not even wrong, but computationally inefficient as it does not
improve the statistical error~(\ref{eq:mperm:errcorr}). 
\subsection{Master equation}
\label{ssec:mperm:master}
Beside ergodicity, another demand for correct statistical sampling
is to ensure that the probability distribution $p({\bf X})$ associated to the desired 
statistical ensemble is independent of time. This can only be
achieved in the simulation, if the \emph{relevant part} of the phase space is sampled sufficiently efficient
to allow for quick convergence towards a stable or, more precisely, stationary
estimate for $p({\bf X})$. In most of the Monte Carlo methods, the simulation
follows a Markov dynamics, i.e., the update of a given conformation ${\bf X}$ to a new
one ${\bf X}'$ is not influenced by the history that led to ${\bf X}$, i.e., the dynamics
does not possess an explicit memory. Such a Markov process can be described by the master equation:
\begin{equation}
\label{eq:mperm:master}
\frac{\Delta p({\bf X})}{\Delta \tau_0}=
\sum\limits_{{\bf X}'} [p({\bf X}')t({\bf X}'\to {\bf X};\Delta \tau_0)-p({\bf X})t({\bf X}\to {\bf X}'; \Delta \tau_0)],
\end{equation}
where $t({\bf X}\to {\bf X}';\Delta \tau_0)$ is the transition probability from ${\bf X}$ 
to ${\bf X}'$ in a single update (or ``time'' step $\Delta \tau_0$). Due to particle conservation,
it satisfies the normalization 
condition $\sum_{{\bf X}'} t({\bf X}\to {\bf X}';\Delta \tau_0) = 1$, i.e., whatever update we perform,
we must end up with a state ${\bf X}'$ which is an element of the conformational space.
The condition $\Delta p({\bf X})/\Delta \tau_0 = 0$ ensures that the ensemble is in a stationary state
if the right-hand side of Eq.~(\ref{eq:mperm:master}) vanishes. Since the stationarity condition
also allows
solutions where the distribution function $p({\bf X})$ dynamically changes on 
cycles which, however, is not the physical situation in a statistical equilibrium ensemble, we
demand more rigorously that the expression in the brackets vanishes. This is called the 
detailed balance condition.\index{detailed balance} Consequently, the ratio of transition rates is given by
\begin{equation}
\label{eq:mperm:detbal}
\frac{t({\bf X}\to {\bf X}'; \Delta \tau_0)}{t({\bf X}'\to {\bf X}; \Delta \tau_0)}=\frac{p({\bf X}')}{p({\bf X})}
\end{equation}
and thus independent of the length of ``time'' step $\Delta \tau_0$, which we,
therefore, omit in the following. 
From this relation, it follows that it is obviously a good idea to construct an efficient
Markov chain Monte Carlo algorithm, i.e., to choose appropriate acceptance probabilities for
the Monte Carlo updates to yield the correct transition probability $t({\bf X}\to {\bf X}')$,
by taking into account the basic microstate probabilities of
the statistical ensemble to be simulated. Markov Monte Carlo simulations in 
the canonical ensemble at fixed temperature $T$, for example, have to satisfy
\begin{equation}
\label{eq:mperm:transcan}
\frac{t({\bf X}\to {\bf X}')}{t({\bf X}'\to {\bf X})}=e^{-\beta \Delta E},
\end{equation}
where $\Delta E=E({\bf X}')-E({\bf X})$ is the energy difference between the new and the old state.
Thus, the transition rate to reach a state ${\bf X}'$, supposed to be
energetically favored if compared with 
the initial state ${\bf X}$, grows exponentially with $\Delta E<0$. ``Climbing the hill'' towards a state
with higher energy ($\Delta E>0$) is, on the other hand, exponentially suppressed. This is in
correspondence with the interpretation of the Markov transition state
theory. Hence, it is possible to study the kinetic behavior (identification of 
free-energy barriers, measuring the height of barriers, estimating transition rates, etc.) of a 
series of processes in equilibrium~-- for example the folding and unfolding behavior of a protein~--
by means of Monte Carlo simulations. To quantify the dynamics of a process, i.e., the explicit time
dependence is, however, less meaningful as the conformational change in a single time step depends
on the move set and does not follow a physical, e.g., Newtonian, dynamics.\footnote{The natural way
to study the time dependence of Newtonian mechanics is typically based on molecular dynamics methods 
which, however, suffer from severe problems to ensure the \emph{correct} statistical sampling at finite
temperatures by using thermostats~\cite{frenkelsmith1,schbj1}. From a more
formal point of view, it is even questionable what ``dynamics''  
shall mean in a thermal system, where even under the same thermodynamic conditions trajectories run typically
differently, due to the ``random'' thermal fluctuations caused by interactions with the huge number 
[${\cal O}$(10$^{\text{23}}$) per mol] of realistically not traceable heat bath particles.}
\subsection{Selection and acceptance probabilities}
\label{ssec:mperm:sel}
In order to correctly satisfy the detailed balance condition~(\ref{eq:mperm:detbal}) 
in a Monte Carlo simulation, we have to take into account that each Monte Carlo
step consists of two parts. First, a Monte Carlo update of the current state
is suggested and second, it has to be decided whether or not to accept it according
to the chosen sampling strategy. In fact, both steps are independent of each other in the sense
that each possible update can be combined with any sampling method. 
Therefore, it is useful to factorize the transition probability $t({\bf X}\to {\bf X}')$
in the selection probability $s({\bf X}\to {\bf X}')$ 
for a desired update from ${\bf X}$ to ${\bf X}'$
and the acceptance probability $a({\bf X}\to {\bf X}')$ for this update:
\begin{equation}
\label{eq:mperm:split}
t({\bf X}\to {\bf X}')=s({\bf X}\to {\bf X}')a({\bf X}\to {\bf X}').
\end{equation}
The acceptance probability is typically used in the form
\begin{equation}
\label{eq:mperm:acc}
a({\bf X}\to {\bf X}')=\min\left(1,\sigma({\bf X},{\bf X'})w({\bf X}\to {\bf X}') \right),
\end{equation}
with the ratio of microstate probabilities
\begin{equation}
\label{eq:mperm:ratio}
w({\bf X}\to {\bf X}')=\frac{p({\bf X'})}{p({\bf X})}
\end{equation}
and the ratio of forward and backward selection probabilities
\begin{equation}
\label{eq:mperm:selratio}
\sigma({\bf X},{\bf X}')=\frac{s({\bf X}'\to {\bf X})}{s({\bf X}\to {\bf X}')}.
\end{equation}
The expression~(\ref{eq:mperm:acc}) for the acceptance probability 
naturally fulfills the detailed-balance condition~(\ref{eq:mperm:detbal}).
The selection ratio $\sigma({\bf X},{\bf X}')$ 
is unity, if the forward and backward selection probabilities
are identical. This is typically the case for ``simple'' local Monte Carlo updates.
If, for example, the update is a translation of a coordinate,
$x'=x+\Delta x$, where $\Delta x\in [-x_0,+x_0]$ is chosen from a uniform 
random distribution, the forward selection for a translation by $\Delta x$
is equally probable to the backward move, i.e., to translate the particle
by $-\Delta x$. This is also valid for rotations about bonds in a molecular system
such as rotations about dihedral angles in a protein. If selection probabilities 
for forward and backward moves differ, the selection rate
is not unity. This is often the case in complex, global updates which comprise
several steps. Then, the determination of the correct selection probabilities can
be difficult and the selection rate has typically to be estimated in test runs first.
To this class of updates belong the biased Gaussian steps~\cite{BGS}, 
where a series of torsional updates of a
few sequential protein backbone 
dihedral angles are performed in order to ensure that the update does not 
drastically change the protein conformation (which would likely be rejected). 

Note that the overall efficiency of a Monte Carlo simulation depends on both,
a model-specific choice of a suitable set of moves and an efficient microstate sampling strategy
based on $w({\bf X}\to {\bf X}')$. 
\subsection{Simple sampling}
\label{ssec:mperm:imp}
The choice of the microstate probabilities $p({\bf X})$ is not necessarily coupled to a certain
physical statistical ensemble. Thus, the simplest choice is a uniform probability $p({\bf X})=1$ 
independently of ensemble-specific microstate properties. Thus also $w({\bf X}\to {\bf X}')=1$
and if the Monte Carlo updates satisfy $\sigma({\bf X},{\bf X}')=1$,
the acceptance probability is trivially also unity, $a({\bf X}\to {\bf X}')=1$, i.e., all
generated Monte Carlo updates are accepted, independently of the type of the update. 
Thus, updates of system degrees of freedom can be performed randomly, where the random numbers
are chosen from a uniform distribution. This method is called \emph{simple sampling}. However,
its applicability is quite limited. Consider, for example, the estimation of the density of states
for a discrete system with this method. After having performed a series of $M$ updates, we will 
have obtained an energetic histogram
$h(E)=M^{-1}\sum_{m=1}^M\delta_{E_m,E}$ which represents an estimate for the
density of states.\index{density of states}
The canonical expectation value of the energy can be estimated by 
$\overline{E}=M^{-1}\sum_{m=1}^M E_m e^{-E_m/k_\text{B}T}=\sum_E E h(E)e^{-E/k_\text{B}T}$.
If the microstates are generated randomly from a uniform distribution, it is obvious that
we will sample the states ${\bf X}$ with an energy $E({\bf X})$ in accordance with their system-specific
frequency or degeneracy. High-frequency states thermodynamically 
dominate in the purely disordered phase. However, near phase transitions towards more ordered phases, 
the density of states drops rapidly~-- typically by many orders of magnitude. The degeneracies
of the lowest-energy states representing the most ordered states are so small that
the thermodynamically most interesting transition region spans even in rather small systems
often hundreds to thousands \emph{orders of magnitude}.\footnote{In order to get an impression of the 
large numbers consider the 2D Ising model of locally interacting spins on a
square lattice which can
only be oriented parallel or antiparallel. For a system with $50\times
50=2500$ spins, the total number of 
spin configurations is thus $2^{2500}\sim 10^{752}$. The degeneracy of the maximally disordered
energetic, paramagnetic is 
of the same order of magnitude. Since the ferromagnetic ground-state degeneracy is 2 (all spins up or all down), i.e., it is of the 
order of 10$^\text{0}$, the density of
states of this rather small system covers far more than 700 orders of magnitude.} 

To bridge a region of 100 orders of magnitude between an ordered and a
disordered phase
by simple sampling would roughly mean to perform about 10$^\text{100}$ updates in order to find
a single ordered state. Assuming that a simple single update would require only a few CPU operations,
it will at least take 1 ns on standard CPU cores. Even under this optimistic assumption, it
would take more than 10$^\text{83}$ years to perform 10$^\text{100}$
updates on a single core! Thus, for studies of complex systems with sufficiently
many degrees
of freedom allowing for cooperativity, simple sampling 
is of very little use. 
\subsection{Metropolis sampling}
\label{ssec:mperm:metro}
\index{Monte Carlo method!Metropolis}
Because of the dominance of a certain restricted space of microstates in ordered phases, 
it is obviously a good idea to primarily concentrate in a simulation
on a precise sampling of the microstates that form the macrostate under given 
external parameters such as, for example, the temperature. The
canonical probability distribution functions clearly show that within the certain
stable phases, only a limited energetic space of microstates is noticeably populated,
whereas the probability densities drop off rapidly in the tails. Thus, an efficient 
sampling of this state space should yield the relevant information within comparatively 
short Markov chain Monte Carlo runs. This strategy is called \emph{importance
  sampling}.\index{importance sampling}

The standard importance sampling variant is the Metropolis method~\cite{metropolis1},
where the algorithmic microstate probability $p({\bf X})$ is identified with the canonical microstate
probability $p({\bf X})\sim e^{-\beta E({\bf X})}$ at the given temperature $T$ ($\beta=1/k_\text{B}T$). Thus, the
acceptance probability~(\ref{eq:mperm:acc}) is governed by the ratio of the 
canonical thermal weights of the microstates:
\begin{equation}
\label{eq:mperm:metro}
w({\bf X}\to{\bf X}')=e^{-\beta [E({\bf X}')-E({\bf X})]}.
\end{equation}
According to Eq.~(\ref{eq:mperm:acc}), a Monte Carlo update from ${\bf X}$ to ${\bf X}'$ 
(assuming $\sigma({\bf X},{\bf X}')=1$) is accepted, if the energy of the new microstate
is lower than before, $E({\bf X}')<E({\bf X})$. If this update would
provoke an increase of energy,
$E({\bf X}')>E({\bf X})$, the conformational change is accepted only with the probability 
$e^{-\beta \Delta E}$, where $\Delta E=E({\bf X}')-E({\bf X})$. 
Technically, in the simulation, a random number 
$r\in[0,1)$ from a uniform
distribution is drawn: If $r\le e^{-\beta \Delta E}$, the move is still accepted, whereas it is
rejected otherwise. Thus, the acceptance probability is exponentially suppressed with $\Delta E$
and the Metropolis simulation yields, at least in principle, a time series which is 
inherently correctly sampled in accordance with the canonical statistics. The arithmetic mean value
of a quantity $O$ over the finite Metropolis time series is already an estimate for the 
canonical expectation value: $\overline{O}=M^{-1}\sum_{m=1}^M O_m\approx \langle O\rangle$. 
In the hypothetical case of an infinitely long simulation ($M\to\infty$), this relation is an exact 
equality, i.e., the deviation is due to the finiteness of the time series only. However, it
is just this restriction to a finite amount of data which limits the quality of Metropolis data.
Because of the canonical sampling, reasonable statistics is only obtained in the 
energetic region which is most dominant for a given temperature, whereas 
in the tails of the canonical distributions
the statistics is rather poor. Thus, there are
three physically particularly interesting cases where Metropolis sampling as standalone method
is little efficient. 

First, for low temperatures, where lowest-energy states dominate, the
widths of the canonical distributions are extremely small and since $\beta\sim 1/T$ is very large, 
energetic ``uphill'' updates are strongly suppressed by the Boltzmann weight $e^{-\beta\Delta E}\to 0$.
That means, once caught in a low-energy state, the simulation freezes and it remains trapped in a low-energy
state for a long period. 

Second, near a second-order phase transition, the standard deviation
$\sigma_E=\sqrt{\langle E^2\rangle -\langle E\rangle^2}$ of the canonical energy 
distribution function gets very large at the critical temperature $T_C$, 
as it corresponds to the maximum (or, in the
thermodynamic limit, the divergence) of the specific $C_V=\sigma_E^2/k_\text{B}T^2$. Thus, a large 
energetic space must precisely be sampled (``critical fluctuations'') which requires high statistics. 
Since in Metropolis dynamics, ``uphill moves'' with $\Delta E>0$ are only accepted with a reasonable rate, if 
at the transition point the ratio $\Delta E/k_\text{B}T_C>0$ is not too large, it can take a long time
to reach a high-energy state if starting from the low-energy end. Since near $T_C$ the correlation
length diverges like $\xi\sim |\tau|^{-\nu}$ [with $\tau=(T-T_C)/T_C$] and the 
correlation time in the Monte Carlo dynamics behaves like $t_\text{corr}\sim |\tau|^{-\nu z}$, 
the dynamic exponent $z$ allows to compare the efficiencies of different algorithms. The larger
the value of $z$, the less efficient is the method. Unfortunately, the standard Metropolis method
turns out to be one of the least efficient methods in sampling critical properties of
systems exhibiting a second-order phase transition.

The third reason is that the Metropolis method does also perform poorly at
first-order phase transitions. In this case, the canonical distribution function is bimodal,
i.e., it exhibits two separate peaks with a highly suppressed energetic region in-between,
since two phases coexist. For
the reasons already outlined, it is extremely unlikely to succeed if trying to
``jump'' from 
the low- to the high-energy phase by means of Metropolis sampling;
it rather would have to explore the valley step by step. Since the energetic region between
the phases is entropically
suppressed~-- the number of possible states the system can assume is simply too small~-- it is thus
quite unlikely that this ``diffusion process'' will lead the system into the high-energy phase, or it
will at least take extremely long. 

However, apart from lowest-energy and phase transition regions, the Metropolis method can 
successfully be employed, often in combination with reweighting techniques.
\section{Reweighting methods}
\subsection{Single-histogram reweighting}
\index{reweighting!single-histogram}
A standard Metropolis simulation is performed at a given temperature, say $T_0$. 
However, it is often desirable to get also quantitative information about the
changes of the thermodynamic behavior at nearby temperatures.
Since Metropolis sampling is not a priori restricted to a limited phase space, 
at least in principle, it is indeed theoretically possible to reweight Metropolis data obtained for
a given temperature $T_0=1/k_\text{B}\beta_0$ to a different one, $T=1/k_\text{B}\beta$.
The idea is to ``divide out'' the Boltzmann factor $e^{-\beta_0 E}$ in the estimates for any quantity 
at the simulation temperature and to multiply it by $e^{-\beta E}$:
\begin{equation}
\label{eq:mperm:shrew}
\langle O\rangle_{T} = 
\frac{\left\langle O e^{-(\beta-\beta_0)E}\right\rangle_{T_0}}{\left\langle e^{-(\beta-\beta_0)E}\right\rangle_{T_0}}
\approx \overline{O}_{T}=
\frac{\sum_{m=1}^MO_me^{-(\beta-\beta_0)E_m}}{\sum_{m=1}^M e^{-(\beta-\beta_0)E_m}},
\end{equation} 
where we have again considered that the MC time series of length $M$ is finite. In practice, the
applicability of this simple reweighting method is rather limited in case the data series was
generated in a single Metropolis run, since the error in the tails
of the simulated canonical histograms rapidly increases with the distance from the peak.
By reweighting, one of the noisy tails will gain the more statistical weight the larger 
the difference between the temperatures $T_0$ and $T$ is. In combination with the generalized-ensemble
methods to be discussed later in this chapter, however, single-histogram reweighting is the only way of
extracting the canonical statistics off the simulated histograms and works perfectly.
\subsection{Multiple-histogram reweighting}
\index{reweighting!multiple-histogram}
From each Metropolis run, an estimate for the density of states\index{density of
states} $g(E)$ can easily be calculated.
Since the histogram measured in a simulation at temperature $T$, $h(E;T)=\sum_{m=1}^M\delta_{E\, E_m}$, is an
estimate for the canonical distribution function $p_{\text{can}}(E;T)\sim g(E)e^{-\beta E}$, the 
estimate for the density of states is obtained by reweighting, $\overline{g}(E)=h(E;T)e^{\beta E}$.
However, since in a ``real'' Metropolis run at the single temperature $T$ accurate data can 
only be obtained in a certain energy interval which
depends on $T$, the estimate $\overline{g}(E)$ is restricted to this typically rather narrow 
energy interval and does by far not cover
the whole energetic region reasonably well. 

Thus, the question is whether the combination of Metropolis data obtained in simulations at 
different temperatures, can yield an improved estimate $\overline{g}(E)$. This is indeed possible
by means of the multiple-histogram reweighting method~\cite{ferr1}, sometimes also called 
``weighted histogram analysis method'' (WHAM)~\cite{ferr2}. Even though the general idea is simple,
the actual implementation is not trivial. The reason is that conventional 
Monte Carlo simulation techniques such as the Metropolis method cannot yield
absolute estimates
for the partition sum $Z(T)=\sum_E g(E)e^{-\beta E}$, i.e., estimates for the 
density of states at different energies $g_i(E)$ and $g_j(E')$ 
can only be related to each other 
if obtained in the
same run, i.e., $i=j$, but not if performed under different conditions. This is not a
problem for the estimation of
mean values or normalized distribution functions at fixed temperatures as long as 
the Metropolis data obtained in the respective temperature threads are used, 
but interpolation to temperatures where no data were explicitly generated, 
is virtually impossible. Also the multiple-histogram reweighting method does not 
solve the problem of getting absolute quantities, 
but at least a ``reference partition function'' is introduced,
which the estimates of the density of states obtained in runs at different simulation 
temperatures can be related to. Thus, interpolating the data between different temperatures
becomes possible.

Basically, the idea is to perform a weighted average of the histograms $h_i(E)$, measured in Monte
Carlo simulations for
different temperatures, i.e., at $\beta_i$ (where $i=1,2,\ldots,I$ indexes the simulation
thread), in order to obtain an estimator for the density of states by
combining the histograms in an optimal way:
\begin{equation}
\label{eq:mperm:whamweight}
\hat{g}(E)=\frac{\sum_i g_i(E) w_i(E)}{\sum_i w_i(E)}. 
\end{equation}
The exact density of states is given by $g(E)=p_{\text{can}}(E;T)Z(T)e^{\beta E}$
and since the normalized histogram $h_i(E)/M_i$ 
obtained in the $i$th simulation thread is an estimator
for the canonical distribution function $p_{\text{can}}(E;T_i)$, the density 
of states is in this thread estimated by
\begin{equation}
\label{eq:mperm:dosest}
g_i(E)=\frac{h_i(E)}{M_i} Z_ie^{\beta_i E},
\end{equation}
where $Z_i$ is the unknown partition function at the $i$th temperature. 
Since in Metropolis simulations
the best-sampled energy region depends on the simulation temperature, the number
of histogram entries for a given energy
will differ from thread to thread. Thus, the data of the thread with high statistics at $E$
should in this interpolation scheme get more weight than histograms with less entries at $E$.
Therefore, the weight shall be controlled by the errors of the individual histograms. A
possibility to determine a set of optimal weights is to reduce the deviation of the estimate
$\hat{g}(E)$ for the density of states from the unknown exact distribution $\langle g\rangle(E)$,
where the symbol $\langle\ldots\rangle$ is used to refer to this quantity as the true
distribution which would have been hypothetically obtained in an infinite number of threads 
(it should not be confused with a statistical ensemble average).
As usual, the ``best'' estimate is the one that minimizes the variance 
$\sigma_{\hat g}^2=\langle (\hat{g}-\langle g\rangle)^2\rangle$. Inserting the 
relation~(\ref{eq:mperm:whamweight}) and minimizing with respect to the weights $w_i$ yields 
a solution 
\begin{equation}
\label{eq:mperm:opthistweight}
w_i=\frac{1}{\sigma_{g_i}^2},
\end{equation} 
where $\sigma_{g_i}^2=\langle(g_i-\langle g_i\rangle)^2 \rangle$ 
is the exact variance of $g_i$ in the $i$th thread. Because of Eq.~(\ref{eq:mperm:dosest}) and the 
fact that $Z_i$ is an energy-independent constant in the $i$th thread, we
can now concentrate on the discussion of the error of the $i$th histogram, since
$\sigma_{g_i}^2=\sigma_{h_i}^2 Z_i^2e^{2\beta_i E}/M_i^2$.

The variance $\sigma_{h_i}^2$ is also an unknown quantity and, in principle,
an estimator for this variance would be needed. This would yield an expression that includes 
the autocorrelation time~\cite{ferr1,ferr2}~-- similar to the discussion below Eq.~(\ref{eq:mperm:errcorr}). 
However, to correctly keep track of the correlations in histogram reweighting is difficult
and thus also the estimation of error propagation is nontrivial. Therefore, we follow the standard 
argument based on the assumption of uncorrelated Monte Carlo dynamics (which is typically not 
perfectly true, of course). The consequence of this idealization will be 
that the weights~(\ref{eq:mperm:opthistweight}) are not necessarily optimal anymore (the
applicability of the method itself is not dependent of the choice of $w_i$, but the error of the 
final histogram will depend on the weights).

In order to determine $\sigma_{h_i}^2$ for uncorrelated data, we only need to calculate the
probability $P(h_i)$ that in the $i$th thread a state with energy $E$ (for simplicity we 
assume that the problem is discrete) is hit $h_i$ times in $M_i$ 
trials, where each hit occurs with the probability $p_\text{hit}$. 
This leads to the binomial distribution with the hit average 
$\langle h_i\rangle=M_i p_\text{hit}$. In the limit of small hit probabilities 
(a reasonable assumption in general if the number of energy bins is large, and, in particular,
for the tails of the histogram) 
the binomial turns into the Poisson distribution 
$P(h_i)\to \langle h_i\rangle^{h_i} e^{-\langle h_i\rangle}/h_i!$
with identical variance and expectation value, $\sigma_{h_i}^2=\langle h_i\rangle$. 
Insertion into Eq.~(\ref{eq:mperm:opthistweight}) yields the weights
\begin{equation}
\label{eq:mperm:optweightB}
w_i(E)=\frac{M_i^2}{\langle h_i\rangle(E)Z_ie^{2\beta_i E}}. 
\end{equation}
Since $\langle h_i\rangle(E)$ is exact, the exact density of states can also be written
as
\begin{equation}
\label{eq:mperm:dosex}
g(E)=\frac{\langle h_i\rangle(E)}{M_i} Z_ie^{\beta_i E}
\end{equation}
which is valid for all threads, i.e., the left-hand side is independent of $i$. This 
enables us to replace $\langle h_i\rangle$ everywhere. Inserting
expression~(\ref{eq:mperm:optweightB}) into Eq.~(\ref{eq:mperm:whamweight}) and utilizing
the relation~(\ref{eq:mperm:dosex}) to replace $\langle h_i\rangle$, we finally end up with
the estimator for the density of states in the form
\begin{equation}
\label{eq:mperm:dosestwham}
\hat{g}(E)=\frac{\sum_{i=1}^I h_i(E)}{\sum_{i=1}^I M_iZ_i^{-1}e^{-\beta_i E }},
\end{equation}
where the unknown partition sum is given by
\begin{equation}
\label{eq:mperm:partfuncwham}
Z_i = \sum_E \hat{g}(E) e^{-\beta_i E},
\end{equation}
i.e., the set of equations~(\ref{eq:mperm:dosestwham}) and~(\ref{eq:mperm:partfuncwham}) must be
solved iteratively.\footnote{Note that for a system with continuous
energy space which is partitioned into bins of width $\Delta E$ 
in the simulation, the right-hand side
of Eq.~(\ref{eq:mperm:partfuncwham}) must still be multiplied by $\Delta E$.} One initializes the recursion with guessed values $Z_i^{(0)}$ for all 
threads, calculates the first estimate $\hat{g}^{(1)}(E)$ using $Z_i^{(0)}$, 
re-inserts this into Eq.~(\ref{eq:mperm:partfuncwham}) to obtain $Z_i^{(1)}$, and continues
until the recursion process has converged close enough to a fixed point.

There is a technical aspect that should be taken into account in an actual calculation. Since the density
of states can even for small systems cover many orders of magnitude and also the Boltzmann
factor can become very large, the application of the recursion relations~(\ref{eq:mperm:dosestwham})
and~(\ref{eq:mperm:partfuncwham}) often results in overflow errors since the floating-point
data types cannot handle these numbers. At this point, it is helpful to change to a logarithmic
representation which however, makes it necessary to think about adding up large numbers
in logarithmic form. Consider the special but important case of 
two positive real numbers $a\ge 0$ and $0\le b\le a$ which are too large to be stored
such that we wish to use their logarithmic representations $a_\text{log}=\log a$ and 
$b_\text{log}=\log b$ instead. Since the result of the addition, $c=a+b$, will also be too
large, we introduce $c_\text{log}=\log c$ as well. The summation is then performed by writing
$c=e^{c_\text{log}}=e^{a_\text{log}}+e^{b_\text{log}}$. Since $a\ge b$ (and thus also
$a_\text{log}\ge b_\text{log}$), it is useful to separate $a$, 
and to rewrite the sum as $e^{c_\text{log}}=e^{a_\text{log}}(1+e^{b_\text{log}-a_\text{log}})$. Taking the 
logarithms yields the desired result, where only the logarithmic representations are needed to
perform the summation: $c_\text{log}=a_\text{log}+\log (1+x)$, where 
$x=b/a =e^{b_\text{log}-a_\text{log}} \in [0,1]$. The upper limit $x=1$ is obviously associated to $a=b$,
whereas the lower limit $x=0$ matters if 
$a\ge 0$, $b=0$.\footnote{At the lower limit, there is a numerical problem, if
$b_\text{log}-a_\text{log}\ll 0$ (or $x=b/a\ll 1$) 
is so small that the \emph{minimum} allowed floating-point
number is underflown by $x$. This typically occurs if $a$ and $b$ differ by many 
tens to thousands orders of magnitude (depending on the floating-point number precision).
In this case, the difference between $c$ and $a$ cannot be resolved, as the 
error in $c_\text{log}=a_\text{log}+{\cal O}(x)$ is smaller than the numerical resolution;
in which case we simply set $c_\text{log}=a_\text{log}$. If this is not acceptable and a higher 
resolution is really needed, non-standard concepts of handling numbers with arbitrary precision
could be an alternative.} Since the logarithm of the density of
states\index{density of states} is proportional
to the microcanonical entropy,\index{microcanonical entropy}\index{entropy!microcanonical} $S(E)\sim \log g(E)$, the logarithmic representation has
even an important physical meaning.
\section{Generalized-ensemble Monte Carlo methods}
\label{sec:mperm:general}
\index{Monte Carlo method!generalized-ensemble}
The Metropolis method is the simplest importance sampling Monte Carlo method and for this reason it is a 
good starting point for the simulation of a complex system. However, it
is also one of the least efficient methods and thus one will often have to face
the question of how to improve the efficiency of the sampling. 
One of the most frequently used ``tricks'' is to employ a modified statistical ensemble
within the simulation run and to reweight the obtained statistics after
the simulation. The simulation is performed in an artificial
\emph{generalized ensemble}. 
\subsection{Replica-exchange Monte Carlo method (parallel tempering)}
\label{ssec:mperm:pt}
\index{Monte Carlo method!parallel tempering}
\index{Monte Carlo method!replica-exchange}
Although not being most efficient, parallel tempering is the most
popular generalized-ensemble method. Advantages are the simple implementation
and parallelization on computer systems with many processor cores. The
Metropolis method samples conformations of the system in a single canonical
ensemble at a fixed temperature, whereas
replica-exchange methods like parallel tempering simulate $I$ ensembles at temperatures
$T_1,T_2,\ldots,T_I$ in parallel (and thus $I$ replicas or instances of the
system)~\cite{sw1,huku1,geyer1}. 
In each of the $I$ temperature threads,
standard Metropolis simulations are performed. A decrease of the autocorrelation
time, i.e., an increase in efficiency, is achieved by exchanging
replicas in neighboring temperature threads after a certain number of Metropolis
steps are performed independently in the individual threads. The
acceptance probability for the exchange of the current conformation $\textbf{X}$ 
at temperature $T_i=1/k_\text{B}\beta_i$ and the conformation
$\textbf{X}'$ at $\beta_j$ is given by 
\begin{equation}  
a({\bf X}\leftrightarrow{\bf X}';\beta_i,\beta_j)=\min(1,\exp\{-(\beta_i-\beta_j)[E(\textbf{X}')-E(\textbf{X})]\}),
\end{equation}
which satisfies the detailed balance condition in this generalized
ensemble.\footnote{In the generalized ensemble composed of two canonical
  ensembles at temperatures $T_i$ and $T_j$, the probability for a state ${\bf
    X}$ at $T_i$ and a state ${\bf X}'$ at $T_j$ reads $p({\bf X}, {\bf X}';T_i,T_j)\sim
  \exp\{-[\beta_iE({\bf X})+\beta_jE({\bf X}')]\}$.} Since the temperature of each
thread is fixed, only a small section of the density of states\index{density of states} can be sampled in
each thread because of the Metropolis limitations. In order to obtain an entire estimate of the density of states,
the pieces obtained in the different threads must be combined in an optimal
way. This is achieved by subsequent multiple-histogram reweighting. The main
advantage of parallel tempering is its high parallelizability. However, the most
efficient selection of the temperature set can be a highly sophisticated
task. One necessary condition for reasonable exchange probabilities is a
sufficiently large overlap of the canonical energy distribution functions in
neighboring ensembles. At very low temperatures, the energy
distribution is typically a sharp-peaked function. Thus, the density of
temperatures must be much higher in the regime of an ordered phase, compared with
high-temperature disordered phases. For this reason, the application of the
replica-exchange method is often not particularly useful for unraveling the
system behavior at very low temperatures or near first-order transitions.
\subsection{Multicanonical sampling}
\label{ssec:mperm:multican}
\begin{figure}
\centering
\includegraphics[scale=1.0]{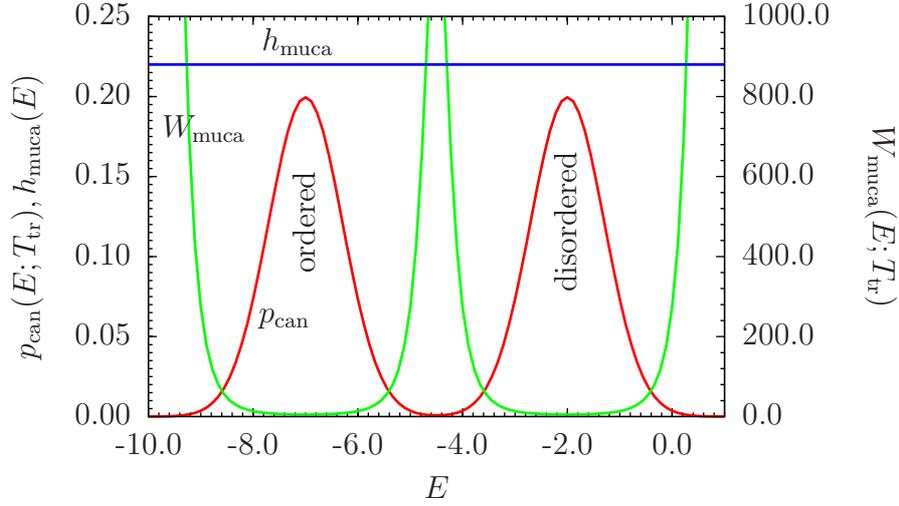}
\caption{\label{fig:muca} Typical scenario of a first-order transition at
  transition temperature $T_\text{tr}$: Ordered
  and disordered phases, represented by the peaked sections of the canonical
  energy distribution $p_\text{can}(E;T_\text{tr})$ at low and high energies,
  are separated by an entropically strongly suppressed energetic region. The
  multicanonical weight function $W_\text{muca}(E;T_\text{tr})$ is chosen in
  such a way that multicanonical sampling provides a random walk in energy
  space, independently of (energetic) free-energy barriers. Thus, the energy distribution
  $h_\text{muca}(E)$ is ideally constant in the multicanonical ensemble.}
\end{figure}

\index{Monte Carlo method!multicanonical}
The powerful multicanonical method~\cite{muca1,muca2,muca3}
makes it possible to scan the whole phase space
within a single simulation with very high accuracy~\cite{baj1}, even if
first-order transitions occur. The principle
idea is to deform the Boltzmann energy 
distribution
$p_{\rm can}(E;T)\propto g(E)\exp(-\beta E)$
in such a way that the
notoriously difficult sampling of the tails is increased and~-- particularly
useful~-- the sampling rate of the entropically 
strongly suppressed lowest-energy conformations is improved. In order to achieve this, the
canonical Boltzmann distribution is modified by the multicanonical weight $W_{\rm muca}(E;T)$
which, in the ideal case, flattens the energy distribution:
\begin{equation}
\label{eq:muca}
W_\text{muca}(E;T)p_{\rm can}(E;T)\sim h_\text{muca}(E) =\text{const}_{E;T},
\end{equation}
where $h_\text{muca}(E)$ denotes the (ideally flat) multicanonical
histogram. By this construction, the multicanonical simulation performs a random
walk in energy space which rapidly decreases the autocorrelation time in
entropically suppressed regions. This is particularly apparent and important in
the phase separation regime at first-order-like transitions, as it is schematically
illustrated in Fig.~\ref{fig:muca}. 

Recalling that the 
simulation temperature $T$ does not
possess any meaning in the multicanonical ensemble as, according to Eq.~(\ref{eq:muca}), the
energy distribution is always constant, independently of temperature.
Actually, it is convenient to set it to infinity in which
case $\lim_{T\to\infty}p_{\rm can}(E;T)\sim g(E)$ and thus
$\lim_{T\to\infty}W_{\rm muca}(E;T)\sim g^{-1}(E)$. Then, the acceptance
probability~(\ref{eq:mperm:acc}) is governed by
\begin{equation}
w({\bf X}\to{\bf X}')=W_\text{muca}(E({\bf X}'))/W_\text{muca}(E({\bf X}))=g(E({\bf X}))/g(E({\bf X}')).
\end{equation}
The weight function can suitably be
parametrized as 
\begin{equation}
W_\text{muca}(E)\sim\exp[-S(E)/k_\text{B}]=\exp\{-\beta(E)[E-F(E)]\},
\end{equation} 
where $S(E)$ is the microcanonical
entropy\index{entropy!microcanonical}\index{microcanonical entropy} $S(E)=k_\text{B}\ln g(E)$. Since
$\beta(E)=\partial S(E)/\partial E$ is the microcanonical thermal energy (with
$\beta(E)=1/k_\text{B}T(E)$, where $T(E)$ is the microcanonical temperature),
the microcanonical free-energy scale $f(E)=\beta(E)F(E)$ and $\beta(E)$ are
related to each other by the differential equation
\begin{equation}
\label{eq:scale}
\frac{\partial f(E)}{\partial E}=\frac{\partial \beta(E)}{\partial E}E.
\end{equation}
Since $\beta(E)$ and $f(E)$ are unknown in the beginning of the simulation, this
relation must be solved recursively. This can be done in an efficient
way~\cite{muca2,muca3,celik1}. If not already being discrete by the model definition, the
energy spectrum must be discretized, i.e., neighboring energy
bins are separated by an energetic step size $\varepsilon$. Thus, for the estimation
of $\beta(E)$ and $f(E)$, the following system of difference equations needs to be solved
recursively ($s(E)=S(E)/k_\text{B}$):
\begin{eqnarray}
\label{eq:recurs}
s^{(n-1)}(E)&=&\ln g^{(n-1)}(E)=-\ln W^{(n-1)}_\text{muca}(E)\nonumber \\
\beta^{(n)}(E)&=&[s^{(n-1)}(E)-s^{(n-1)}(E-\varepsilon)]/\varepsilon\nonumber\\
f^{(n)}(E)&=&f^{(n)}(E-\varepsilon)+[\beta^{(n)}(E)-\beta^{(n)}(E-\varepsilon)](E-\varepsilon)\\
s^{(n)}(E)&=&\beta^{(n)}(E)E-f^{(n)}(E)\nonumber\\
W^{(n)}_\text{muca}(E)&=&\exp[-s^{(n)}(E)].\nonumber
\end{eqnarray}
The superscript $(n)$ refers to the index of the iteration. If no better initial
guess is available, one typically sets $g^{(0)}(E)=1$ in the beginning, implying
$s^{(0)}(E)=W^{(0)}_\text{muca}(E)=0$. The zeroth iteration thus corresponds to
a Metropolis run at infinite temperature, yielding the first estimate for the
multicanonical weight function $W^{(1)}_\text{muca}(E)$, which is used to
initiate the second recursion, etc. The
recursion procedure based on Eq.~(\ref{eq:recurs}) can be stopped after $I$ recursions, if the weight
function has sufficiently converged. The number of necessary recursions and also
the number of sweeps to be performed within each recursion is model
dependent. Since the sampled energy space increases from recursion to recursion
and the effective statistics of the histogram in each energy bin depends on the
number of sweeps, it is a good idea to increase the number of sweeps successively
from
recursion to recursion. Since the energy histogram should be ``flat'' after the
simulation run at a certain recursion level, an alternative way to control the
length of the run is based on a flatness criterion. If, for example, minimum and
maximum value of the histogram deviate from the mean histogram value by less
than 20\%, the run is stopped.

Finally, after the best possible estimate for the
multicanonical weight function is obtained, a long multicanonical production run
is performed, including all measurements of quantities of interest. From the
multicanonical trajectory, the estimate of the canonical expectation value of a quantity $O$ is
then obtained at any (canonical) temperature
$T$ by:
\begin{equation}
\overline{O}_T=\frac{\sum_tO(\textbf{X}_t)W^{-1}_\text{muca}(E(\textbf{X}_t)
  e^{-E(\textbf{X}_t)/k_\text{B}T}}{\sum_t W^{-1}_\text{muca}(E(\textbf{X}_t)
  e^{-E(\textbf{X}_t)/k_\text{B}T}}.
\end{equation}
Since the accuracy of multicanonical sampling is independent of the canonical
temperature and represents a random walk in the entire energy space, the
application of reweighting procedures is lossless. This is a great advantage of
the multicanonical method, compared with Metropolis Monte Carlo
simulations. Virtually, a multicanonical simulation samples the system behavior
at \emph{all} temperatures simultaneously, or, in other words, the direct
estimation of the density of states is another advantage, because
multiple-histogram
reweighting is not needed for this (in contrast to replica-exchange methods). 
\subsection{Wang-Landau method}
\label{ssec:mperm:wl}
\index{Monte Carlo method!Wang-Landau}
In multicanonical simulations, the weight functions are updated after each
iteration, i.e., the weight and thus the current estimate of the density of
states\index{density of states} are kept constant at a given recursion level. For this reason, the
precise estimation of the multicanonical weights in combination with the
recursion scheme~(\ref{eq:recurs}) can be a complex and not very efficient
procedure. In the method introduced by Wang and Landau~\cite{wl1}, the
density of states estimate is changed by a so-called modification factor
$\alpha$ after
each sweep, $g(E)\to \alpha^{(n)} g(E)$, where $\alpha^{(n)}> 1$ is kept constant
in the $n$th recursion, but it is reduced from iteration to iteration. A frequently
used ad hoc modification factor is given by
$\alpha^{(n)}=\sqrt{\alpha^{(n-1)}}=(\alpha^{(0)})^{1/2^n}$, $n=1,2,\ldots,I$, where
often $\alpha^{(0)}=e^1=2.718\ldots$ is chosen. The acceptance
probability and histogram flatness criteria are the same as in multicanonical sampling.

Since the dynamic modification of the density of states in the running simulation
violates the detailed balance condition~(\ref{eq:mperm:detbal}), the advantage of
the high-speed scan of the energy space is paid by a systematic error. However,
since the modification factor is reduced with increasing iteration level until
it is very small (the iteration process is typically stopped if $\alpha<
1.0+10^{-8}$), the simulation dynamics is supposed to sample the phase space
according to the stationary
solution of the master equation such that detailed balance is (almost) satisfied. 
Since it is difficult to keep this convergence under control, the optimal
method is to use the Wang-Landau method for a very efficient generation of the multicanonical
weights, followed by a long multicanonical production run (i.e., at exactly
$\alpha=1$) to obtain the statistical data.
\section{Summary}
Monte Carlo computer simulations are virtually the only way to analyze the
thermodynamic behavior of a system in a precise way. However, the various
existing methods exhibit extreme differences in their efficiency, depending on
model details and relevant questions. The original standard method, Metropolis
Monte Carlo, which provides only reliable statistical information at a given
(not too low) temperature has meanwhile been replaced by more sophisticated
methods which are typically far more efficient (the differences in time scales
can be compared with the age of the universe). However, none of the methods
yields automatically accurate results, i.e., a system-specific adaptation and
control is always needed. Thus, as in any good experiment, the most important
part of the data analysis is statistical error estimation.





\newpage


\end{document}